\documentclass{article}

\usepackage{PRIMEarxiv}

\usepackage[utf8]{inputenc} % allow utf-8 input
\usepackage[T1]{fontenc}    % use 8-bit T1 fonts
\usepackage{hyperref}       % hyperlinks
\usepackage{url}            % simple URL typesetting
\usepackage{booktabs}       % professional-quality tables
\usepackage{amsfonts}       % blackboard math symbols
\usepackage{nicefrac}       % compact symbols for 1/2, etc.
\usepackage{microtype}      % microtypography
\usepackage{lipsum}
\usepackage{fancyhdr}       % header
\usepackage{graphicx}       % graphics
\graphicspath{{media/}}     % organize your images and other figures under media/ folder
\usepackage{subcaption}
\usepackage{amsmath}

\usepackage{amssymb}
\usepackage{subcaption}

\usepackage{floatrow}
\usepackage{lineno}
\title{Deformable multi-modal image registration for the correlation between optical measurements and histology images
}

\author{
 Lianne Feenstra, Maud Lambregts, Theo J.M Ruers and Behdad Dashtbozorg \\
  Image-Guided Surgery \\
  Netherlands Cancer Institute \\
  Amsterdam\\
  \texttt{\{l.feenstra, t.ruers, b.dasht.bozorg\}@nki.nl} \\
}
\begin{document}
\maketitle
%%%%%%%%%%%%%%%%%%% Abstract %%%%%%%%%%%%%%%%

\begin{abstract}
The correlation of optical measurements with a correct pathology label is often hampered by imprecise registration caused by deformations in histology images. This study explores an automated multi-modal image registration technique utilizing deep learning principles to align snapshot breast specimen images with corresponding histology images. The input images, acquired through different modalities, present challenges due to variations in intensities and structural visibility, making linear assumptions inappropriate. An unsupervised and supervised learning approach, based on the VoxelMorph model, was explored, making use of a dataset with manually registered images used as ground truth. Evaluation metrics, including Dice scores and mutual information, reveal that the unsupervised model outperforms the supervised (and manual approach) significantly, achieving superior image alignment. This automated registration approach holds promise for improving the validation of optical technologies by minimizing human errors and inconsistencies associated with manual registration. 
\end{abstract}

%%%%%%%%%%%%%%%%%%%%%%%%%%  Introduction  %%%%%%%%%%%%%%%%%%%%%%%%%%
\section{Introduction}
Optical technologies revolutionized the field of oncologic surgery in recent years by providing non-invasive and innovative ways to monitor the assessment of resection margins during surgical procedures. By providing real-time visualization of tissue characteristics, optical technologies, such as Diffuse Reflectance Spectroscopy (DRS) \cite{deBoer2018TowardsSurgery, Veluponnar2023DiffuseFibers}, Fluorescence lifetime imaging (FlIm) \cite{Alfonso-Garcia2021MesoscopicAccess}, and Hyperspectral imaging \cite{Jong2022DiscriminatingImaging, Kho2021FeasibilitySpecimen}, can help to assess if all cancerous tissue is removed while minimizing damage to surrounding healthy tissue. This will lower the number of positive resection margins and thereby reduces the need for additional treatments such as surgical re-excision or radiotherapy. However, the accuracy and reliability of these technologies are crucial to ensure the safety and efficacy during surgical procedures. Therefore, validation of optical technologies is essential to establish their clinical significance and provide healthcare professionals with the confidence to use them in their practice. This involves assessing the clinical outcomes of these technologies with the current ground truth \cite{Wilson2018ChallengesImaging}. 

Ground truth validation of optical tissue measurements refers to the process of comparing the acquired measurements with the gold standard histopathological analysis of tissue samples. This is provided by hematoxylin and eosin (H\&E) stained tissue sections, from which the measured tissue can be defined microscopically \cite{Wells2007ValidationView}. To ensure an accurate correlation between the performed measurement locations and their corresponding H\&E sections, it is essential to track the locations in order to locate them back microscopically [other paper]. Subsequently, a registration between a specimen snapshot image (with tracked measurement locations) and corresponding H\&E sections (microscopic histology image) is necessary, enabling validation of the measured tissue types against ground truth. This process is crucial for establishing the reliability and reproducibility of optical techniques used for tissue diagnosis, and especially for the accurate development of tissue classification algorithms. 

However, the registration of a specimen snapshot image with the corresponding histology image encounters some challenges. The histopathological processing of tissue specimens which involves steps such as fixation, dehydration, clearing, embedding, and cutting causes tissue deformation in H\&E sections. This deformation may include shrinkage, stretching, compression, tearing, and even loss of tissue \cite{PrittTheBBB}. Also, other factors have an influence on deformation. For example, breast tissue, due to the presence of fat, is more prone to shrinkage or compression during processing compared to muscle tissue. Similarly, the size and thickness of tissue sections can affect the degree of tissue deformation, with thicker sections more susceptible to distortion. Besides, over-staining or prolonged staining influences the amount of deformation, while under-staining may lead to poor visualization of tissue structures \cite{McinnesArtefactsHistopathology}.When validating optical measurements, it is crucial to take these tissue deformations into account \cite{Unger2018MethodAssessment}. Especially, when using labeled optical data for the development of machine learning models, where incorrectly labeled data will influence the performance of tissue classification and ultimately impacting clinical outcomes.

However, labeling and validating optical measurements with histopathology, is a subject that has received limited attention in the existing research literature \cite{Naranjo2014StainedDetection,Halicek2018DeformableDemons,Lu2014HyperspectralAccess}. Where in some studies, the aspect of tissue deformation is not even taken into consideration \cite{Phipps2018AutomatedImaging, Laughney2012ScatterAssessment}. In the method proposed by de Boer \textit{et al.} a manual point-based deformable registration between specimen snapshot and H\&E sections by looking for identical landmarks in both images \cite{DeBoer2019MethodDeformations}. The proposed method addressed the need for a deformable registration method when labeling and validating optical measurements since it demonstrated a higher accuracy compared to a method that neglected such deformations. However, a manual point-based registration lacks subjectivity since the identification of corresponding landmarks can vary between different users. This can lead to inconsistent results and reduce the reliability and accuracy of the registration. Also, this labor-intensive process can be time-consuming, particularly when dealing with large datasets. This approach is also not suitable for images with a limited number of paired distinguishable landmarks, which is often the case when registering multi-modal images. 

In general, multi-modal image registration is a complex task that encounters various difficulties. One major challenge arises from the differences in intensity and contrast between images acquired in different modalities. These variations make it challenging to establish accurate paid landmarks between images. Another obstacle is the structural dissimilarity between modalities, resulting in differences in shape, size, and appearance of visible corresponding structures. Non-linear deformations and limited overlapping information further complicate the registration process \cite{Alam2019ChallengesRegistration}. When registering specimen snapshots with histology images, microscopic artifacts such as tears, holes and loss of tissue introduce additional complexities to accomplish an accurate registration. Overcoming these challenges involves the development of advanced algorithms capable of handling variations in intensity, contrast, shape, as well deformations.

Automating the registration process, by using advanced algorithms and computational techniques, shows potential to address current limitations and enhance the overall registration efficiency and thereby improve the accuracy of validating optical technologies \cite{Pei2023ADirections}. Therefore, the purpose of this study is to develop a multi-modal image registration model which is also able to compensate for tissue deformations automatically. The proposed approach is based on the VoxelMorph model which has been adapted to the needs of a multi-modal 2D registration between specimen snapshot images and microscopic histology images.% Thereby, evaluating their performance by comparing the accuracy with previous developed manual approach. 
With this deformable multi-modal image registration model we aim to achieve a faster and more precise method for labeling optical measurements with a ground truth which can lead to a more accurate development of tissue classification algorithms, enhancing their practical use in clinical settings.   

%The novel contributions of this paper can be summarized as follows:
%The remainder of this paper is organized as follows: 

%%%%%%%%%%%%%%%%%%%%%%%%%%  Method and Materials  %%%%%%%%%%%%%%%%%%%%%%%%%%
\section{Material and Methods}
\label{materials and methods}

\subsection{Materials}
\label{materials}
\begin{figure}[h!]
\captionsetup[subfigure]{justification=centering}
        \centering
    \begin{subfigure}{0.23\textwidth}
        \centering
        \includegraphics[width=0.95\textwidth]{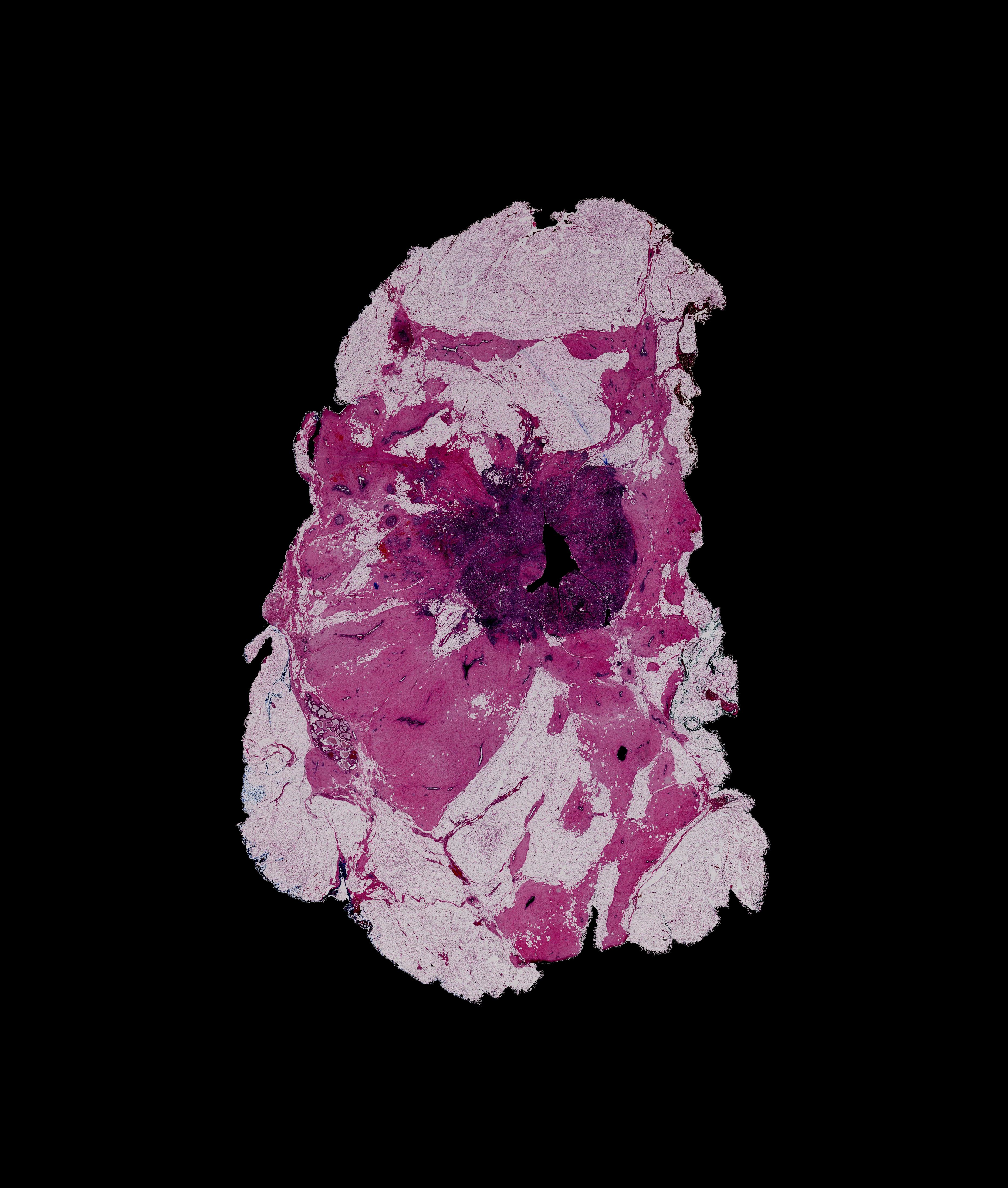}
        \caption{}
        \label{}
    \end{subfigure}
    % \hfill
    \begin{subfigure}{0.23\textwidth}
        \centering
        \includegraphics[width=0.95\textwidth]{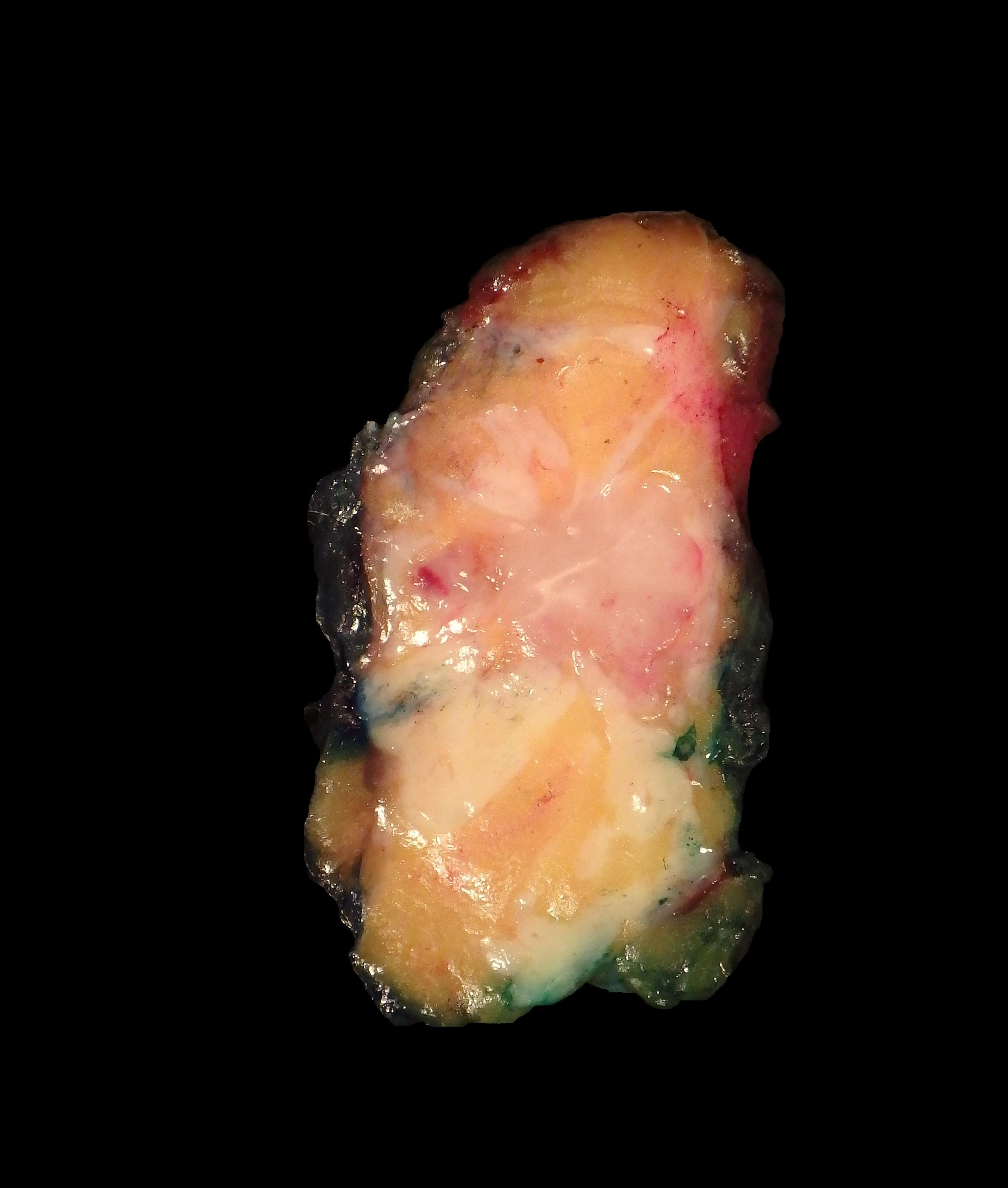}
        \caption{}
        \label{}
    \end{subfigure}
        \begin{subfigure}{0.23\textwidth}
        \centering
        \includegraphics[width=0.95\textwidth]{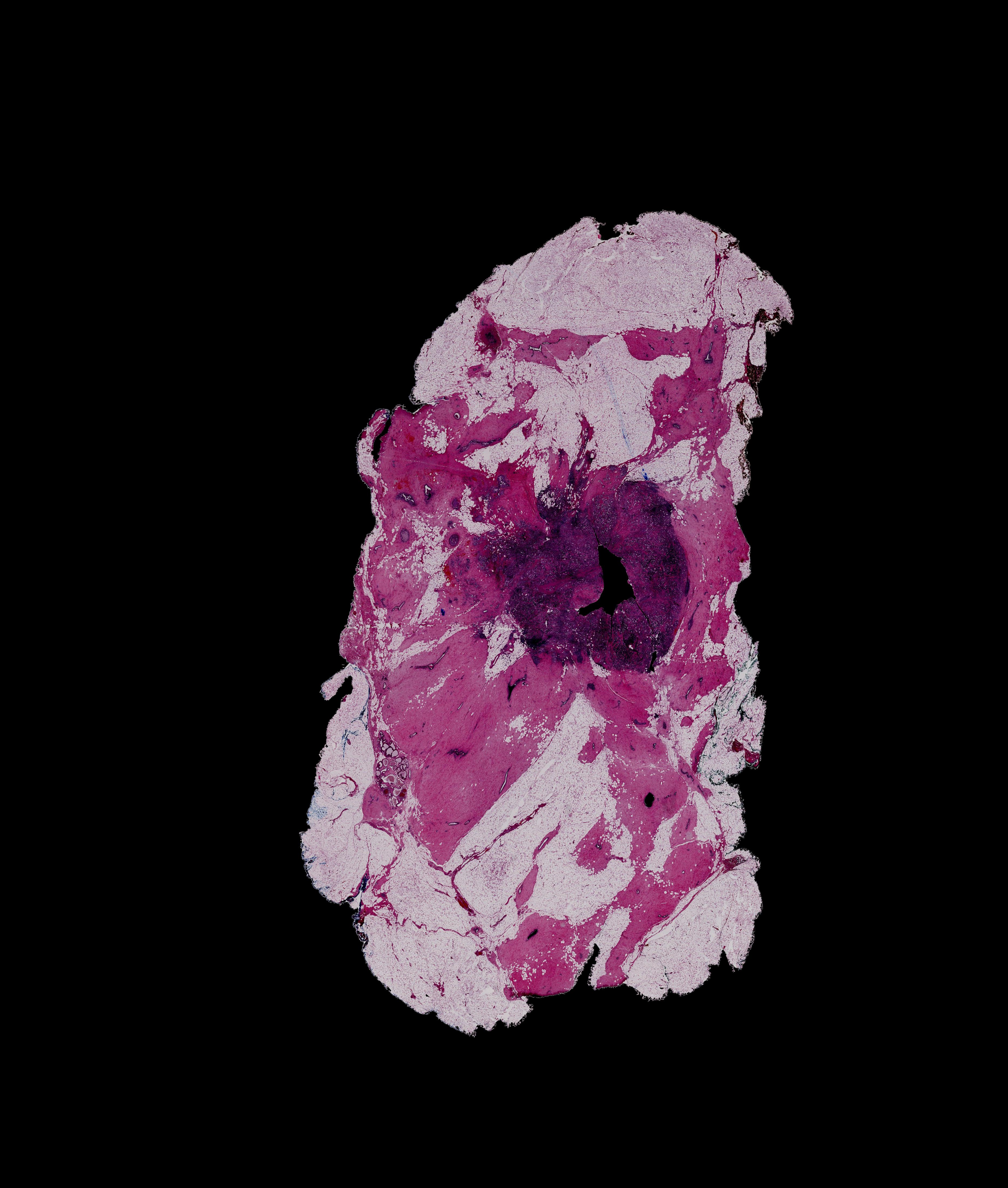}
        \caption{}
        \label{}
    \end{subfigure}
    % \hfill
    \caption{Dataset example: (a) microscopic histology image, (b) snapshot image of the corresponding breast tissue slice captured with a camera, and (c) the manual registered histology image.}
    \label{dataset}
\end{figure}

The dataset used in this study consists of 113 breast tissue slices, each of which comprises three distinct images: a snapshot image of the breast tissue slice captured with a camera, a corresponding microscopic Hematoxylin and Eosin (H\&E) histology image, and a manual registered histology image. Example images of one tissue slice are demonstrated in Figure \ref{dataset}. The manual registration is performed by manual selection of approximately 60 paired control points followed by a deformable registration using a nonrigid local weighted mean transformation, as described in de Boer \textit{et al.}\cite{DeBoer2019MethodDeformations}. Despite the possibility of some misalignment and registration errors, we regard the obtained manual registered histology image in this study as a ground truth image.

\subsection{Method}
\label{method}
In this paper, a multi-modal image registration technique was developed which is capable of automatically addressing tissue deformations, leading to a precise alignment of a snapshot image of a breast tissue slice and the corresponding histology image. 
The proposed multi-modal image registration methods are based on the VoxelMorph medical image registration framework, which will be explained in the following subsection. In this study, we intended to expand the application of VoxelMorph, as the input images for this study were acquired from different modalities.
The development of this multi-modal deformable image registration involves a series of steps, beginning with dataset preparation. This is followed by two different deep-learning approaches for multi-modal image registration using unsupervised and supervised learning models, which will be evaluated separately. 

\subsubsection{VoxelMorph implementation}
\label{VoxelMorph implementation}
The VoxelMorph framework uses an unsupervised deep-learning model for deformable medical image registration. The model is initially designed to work with 3D medical image volumes, such as MRI or CT scans, and can register two volumes of different shapes and sizes without requiring any explicit ground truth registration fields or anatomical landmarks \cite{Balakrishnan2019VoxelMorph:Registration}. 
The architecture of VoxelMorph is based on a deep convolutional neural network ($g_\theta(F, M)$), similar to the UNet model \cite{Ronneberger2015U-Net:Segmentation}. The network uses a moving image $M$ and fixed image $F$ as input and computes a dense displacement field ($\varphi$) based on a set of learnable parameters $\theta$. The network uses this set of parameters to compute the kernels of the convolutional layers, and employs a spatial transformation function to evaluate the similarity between the predicted image ($M(\varphi)$) and the fixed image ($F$). This allows the model to refine its estimation of the optimal spatial transformation function and update its parameters \cite{Jaderberg2015SpatialNetworks}. The generated dense displacement field (DDF) represents the displacement of each pixel in the moving image relative to the corresponding pixel in the fixed image. This dense map of vectors, with the same dimensions as the moving image, describes the spatial transformation required to align $M$ with $F$ which results in the predicted image ($M(\varphi)$). 

The network is trained on an image dataset by minimizing the loss function ($L$) in each epoch, as described in \ref{equation1}. 
\begin{equation}
    \label{equation1}
    L(F,M,\varphi) = L_{sim}(F,M,(\varphi)) + \lambda L_{smooth}(\varphi)
\end{equation}
The loss function $L$ consists of two components: $L_{sim}$ penalizes the difference between the fixed ($F$) and moving ($M$) image, and $L_{smooth}$ is a regularization on the dense deformation field ($\varphi$). The regularization parameter ($\lambda$) defines the weights of the two components. The VoxelMorph network is compatible with any differentiable loss function $L$ \cite{Balakrishnan2018AnRegistration}.

\subsubsection{Data preparation}
\begin{figure}[!h]
   \centering
    \includegraphics[width=0.80\textwidth]{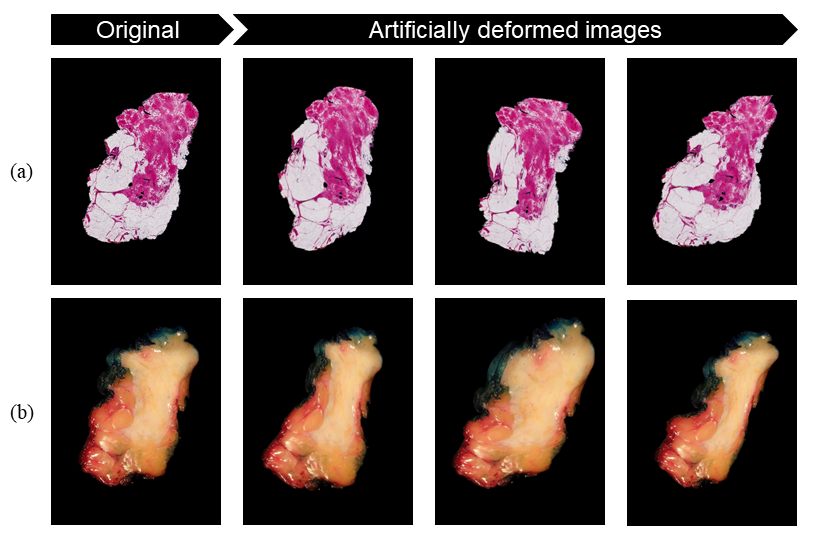}
   \caption{Examples of synthetic deformation applied to histology (a) and specimen snapshot images (b). In this study, artificially deformed histology images were used for training the unsupervised model, whereas artificially deformed specimen snapshot images were used to train the supervised model.}
    \label{deformations}
\end{figure}

Data augmentation is used to increase the number of images in the training set, as well as variations in deformations, which could improve the learning process of the network. Synthetic deformed images were generated from the existing dataset to simulate more deformation variations that occur during the pathology process. 
The augmented images were generated using randomly created dense displacement fields (DDF), in which a number between -1 and +1 was generated for every pixel in both the $x$- and $y$- directions, resulting in displacement fields $\Delta x$ and $\Delta y$. The $\Delta x$ and $\Delta y$ displacement are then convolved with a Gaussian filter with defined filter size $F$ and standard deviation $\sigma$. Here, $\sigma$ is serving as the elasticity coefficient. A scaling factor range $\alpha$ is then applied to the DDF to control the intensity of the deformation \cite{Simard2003BestAnalysis}. The deformation variables ($\sigma$, $\alpha$, $F$) are chosen randomly within a specified range based on the chosen level of deformation intensity resulting in a total of 565 deformed specimen snapshots and histology images. Figure \ref{deformations} illustrates some examples of artificial deformations for different deformation intensity levels.
%This dataset will be split into three subsets, whereas, 360 paired images will be allocated for the training set, 90 paired images for the validation set, and 115 paired images for the test set. 

\begin{figure}[!h]
\captionsetup[subfigure]{justification=centering}
        \centering
    \begin{subfigure}{0.19\textwidth}
        \centering
        \includegraphics[width=0.95\textwidth]{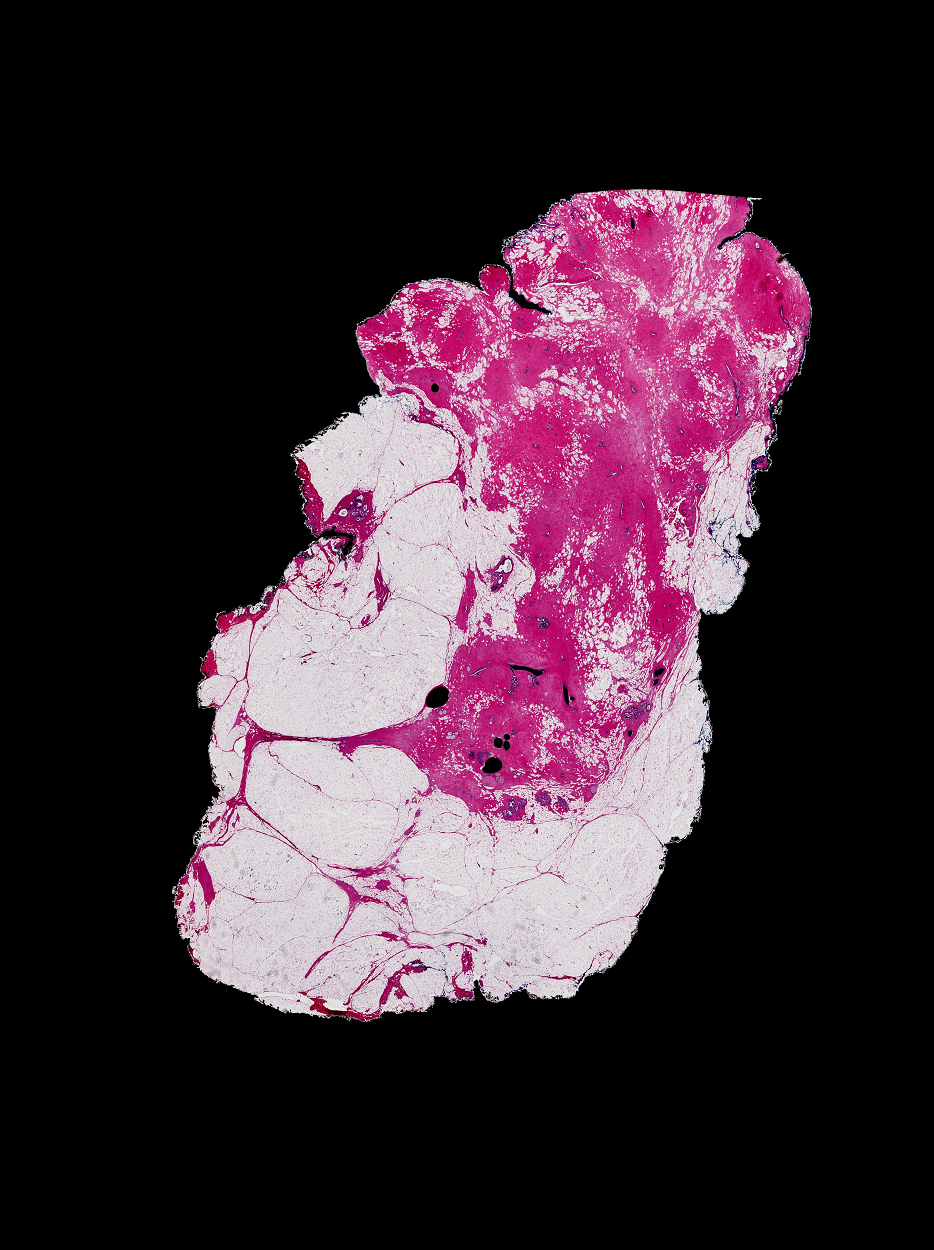}
        \caption{}
        \label{fig:}
    \end{subfigure}
    % \hfill
    \begin{subfigure}{0.19\textwidth}
        \centering
        \includegraphics[width=0.95\textwidth]{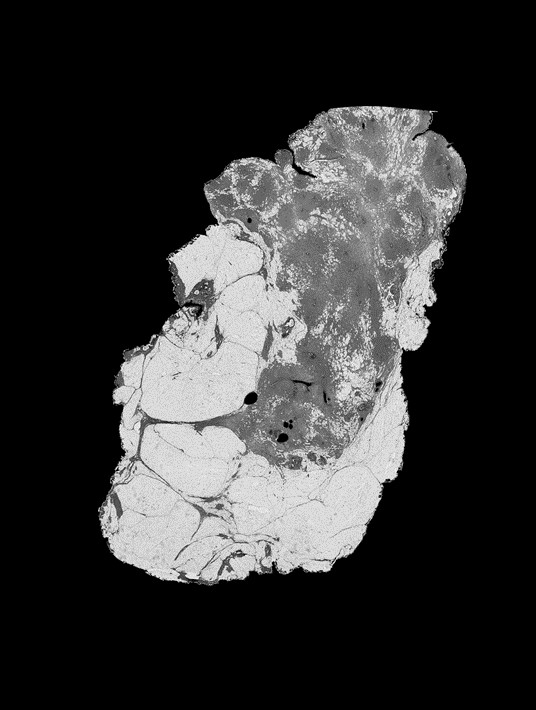}
        \caption{}
        \label{fig:}
    \end{subfigure}
        \begin{subfigure}{0.19\textwidth}
        \centering
        \includegraphics[width=0.95\textwidth]{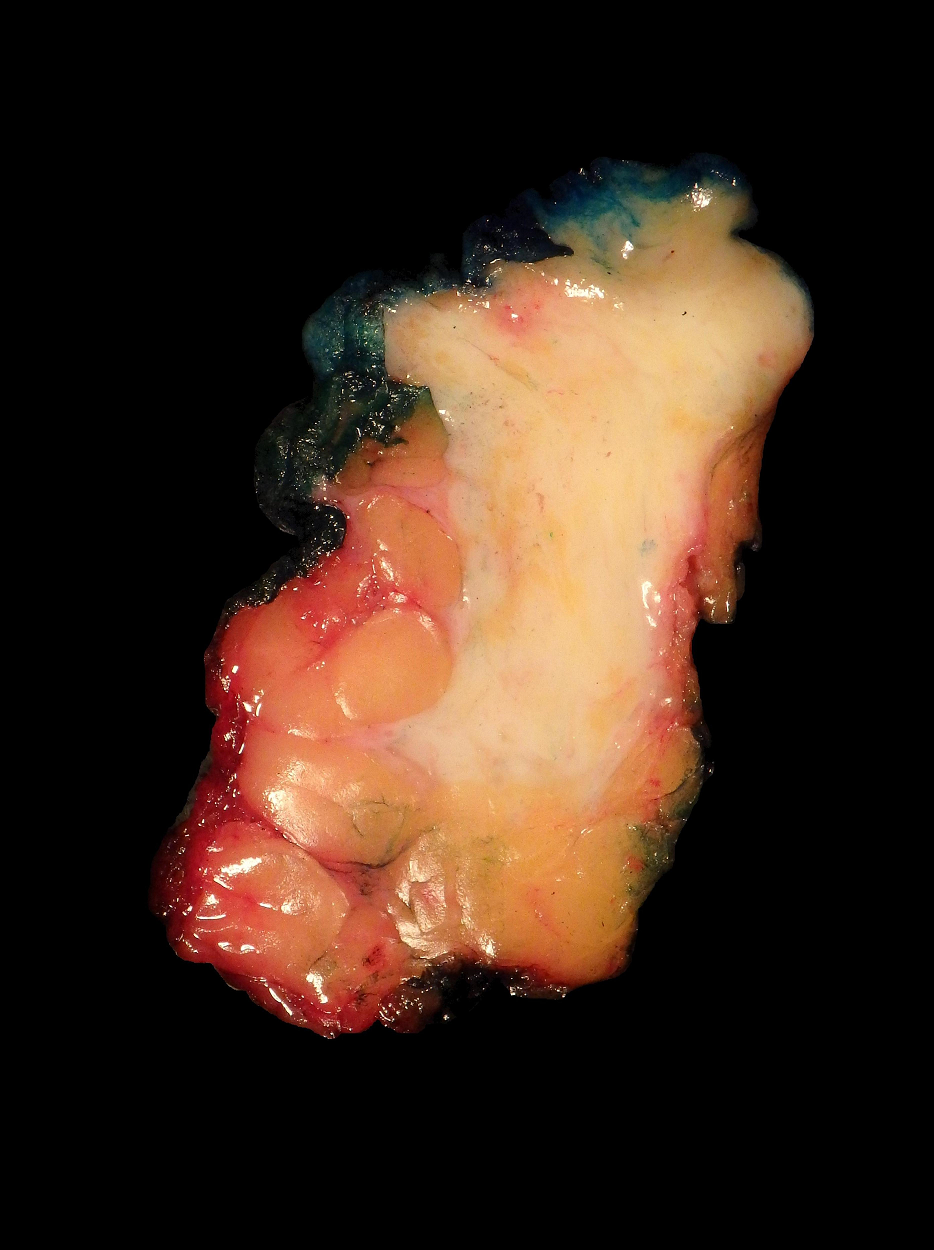}
        \caption{}
        \label{}
    \end{subfigure}
    \begin{subfigure}{0.19\textwidth}
        \centering
        \includegraphics[width=0.95\textwidth]{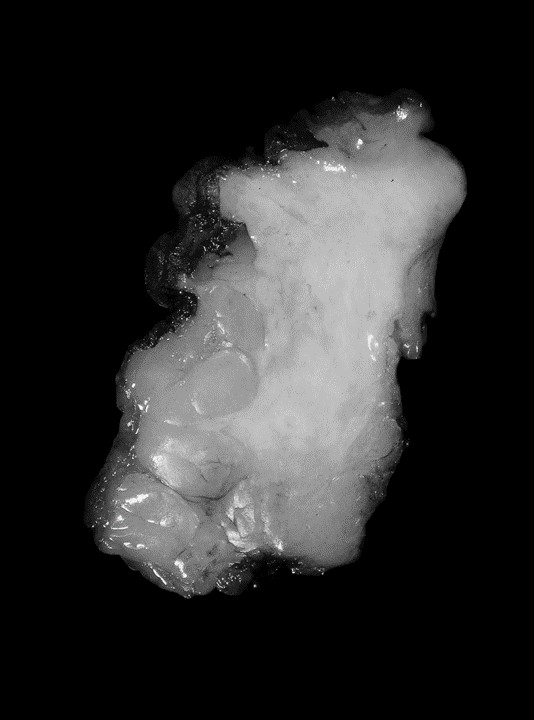}
        \caption{}
        \label{}
    \end{subfigure}
    \begin{subfigure}{0.19\textwidth}
        \centering
        \includegraphics[width=0.95\textwidth]{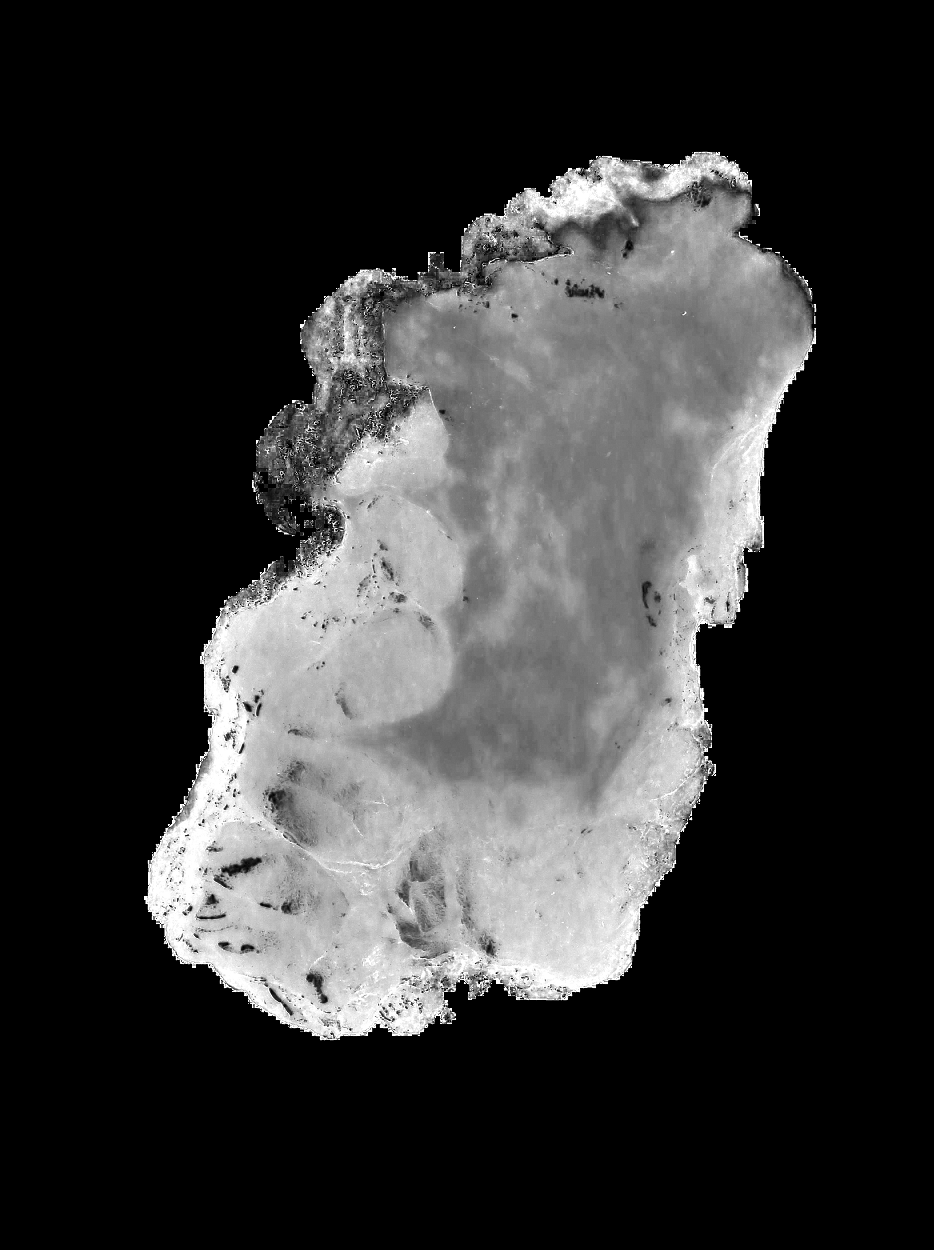}
        \caption{}
        \label{}
    \end{subfigure}
    % \hfill
    \caption{Example of the preprocessing of the used input images: (a) Original RGB histology image (b) grayscale converted histology image (c) original RGB specimen snapshot image (d) grayscale converted specimen snapshot image (e) converted specimen snapshot image using saturation values only.}
    \label{preprocessing}
\end{figure}

Since the input of the VoxelMorph network consists of a 2-channel image representing fixed $F$ and moving $M$ images, it is required to convert both RGB specimen snapshot and histology images to one-channel grayscale images. The weighted average of all color channels (red, green, blue) was used and determined the final grayscale representation, allowing for selective emphasis on certain colors and structures in the histology image. To ensure a similar intensity level between both images, the specimen snapshot images were converted to grayscale by using saturation values only (Figure \ref{preprocessing}). This conversion method enhances the visual correspondence between connective and tumor tissue and is hypothesized to improve the performance of the model. At last, the computational effort and training time for the networks were reduced by resizing the histology and snapshot input images to 256$\times$192 pixels. 

\subsubsection{ Unsupervised learning model}
\label{unsupervised model}
\begin{figure}[!h]
    \centering
    \includegraphics[width=0.95\textwidth]{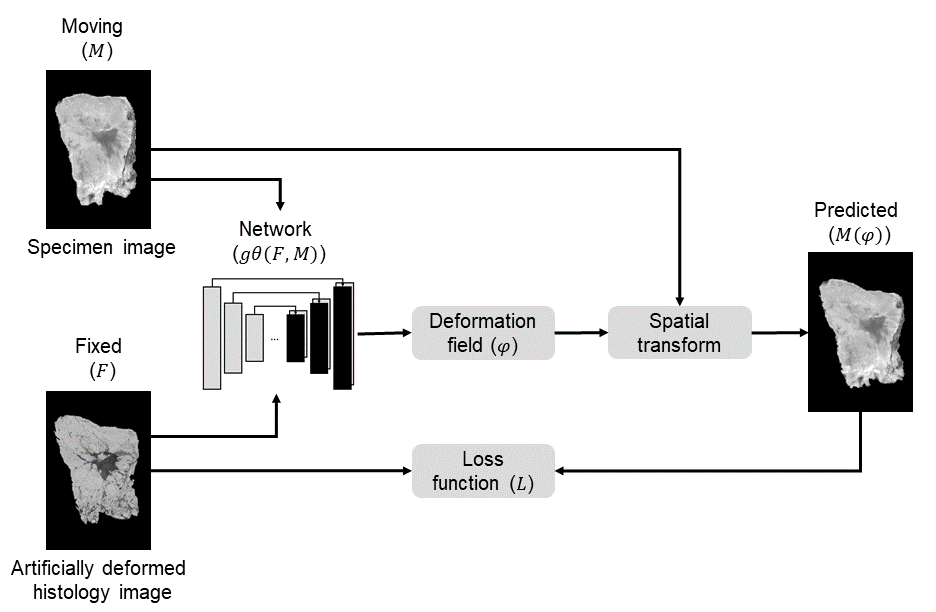}
    \caption{Unsupervised learning model: Specimen snapshot image ($M$) and the artificially deformed histology image ($F$) are used as input images for the unsupervised deep convolutional neural network ($g_\theta(F, M)$). Mutual information is used as loss function ($L$). The network outputs a dense displacement field (DDF($\varphi$)) which defines the mapping from moving image coordinates to the fixed image and is used to register $M$ with $F$. This results in predicted image $(M(\varphi)$).}
    \label{unsupervised}
\end{figure}

In the unsupervised learning approach (Figure \ref{unsupervised}), the input to the model comprises pairs of synthetic deformed histology images ($F$), which imitate the deformations during the pathology process, together with the snapshot specimen images ($M$). The trained network is similar to the original VoxelMorph model (as explained in Section \ref{VoxelMorph implementation}), and entails training $g_\theta(F, M)$ using the input images $F$ and $M$ to compute optimal learnable parameters $\theta$. $M$ will be transformed using the estimated DDF ($\varphi$) in combination with a generated spatial transformer function, resulting in the predicted image ($M(\varphi)$). 
\bigskip 

In this study, the input images were acquired through different modalities. Considering the variations in intensities and structure visibility in both images, it cannot be assumed that the relationship between intensities in these two images is linear. Therefore, mutual information is used as a loss function ($L$) to quantify the statistical dependence between the two images based on their joint distribution. Mutual information (MI) measures the amount of information shared between the two images. In the context of the developed model, the goal is to find a deformation field that maximizes the MI between the two input images. To compute the MI, a histogram-based mutual information ($HMI$) was used, which computes the probability distribution of the intensity values between the two input images, and estimates the joint probability distribution between their histograms \cite{Pluim2003Mutual-information-basedSurvey}. Specifically, $HMI$ is defined as:

\begin{equation}
\label{equation2}
    \text{HMI}(F,M) = \sum_{i,j} p(i,j) \log \frac {p(i,j)}{p(i)p(j)}
\end{equation}

Where, $p(i,j)$ is the joint probability of the intensity values $i$ and $j$ in images $F$ and $M$, and $p(i)$ and $p(j)$ are the marginal probabilities of intensity values $i$ and $j$ in images $F$ and $M$, respectively. By replacing $L_{sim}$ in Equation \ref{equation1} with $HMI$ (Equation \ref{equation2}), the predicted image ($M(\varphi)$) was optimized by maximizing the mutual information between $F$ and $M$.

\subsubsection{Supervised learning model}
\label{supervised model}
\begin{figure}[!h]
   \centering
    \includegraphics[width=0.95\textwidth]{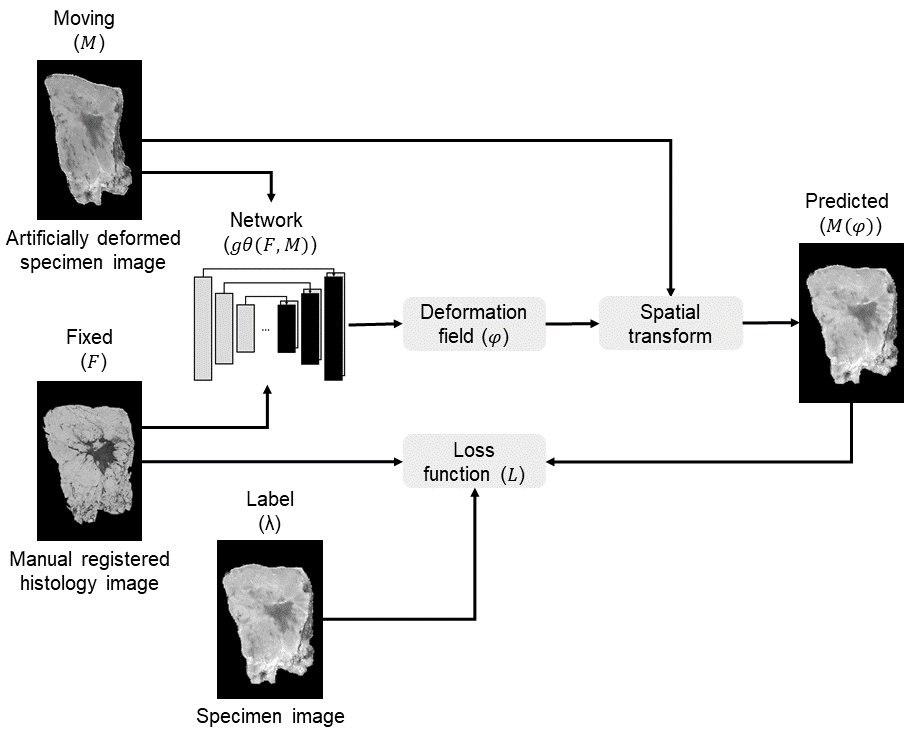}
   \caption{Supervised learning model: Artificially deformed specimen snapshot image ($M$) and the manual registered histology image ($F$) are used as input images for the supervised deep convolutional neural network ($g_\theta(F, M)$). The specimen snapshot images are used as ground truth label ($gamma$). Mean squared error is used as loss function ($L$). The network outputs a dense displacement field (DDF($\varphi$)) which defines the mapping from moving image coordinates to the fixed image and is used to register $M$ with $F$. This results in predicted image $(M(\varphi)$).}
    \label{supervised}
\end{figure}

The VoxelMorph model was originally designed for unsupervised image registration, allowing it to learn without the need for ground truth labels. However, in this study, our dataset consists of manually registered histology images, which can be utilized to train the model in a supervised approach. To achieve this, the moving images ($M$) consist of artificially deformed snapshot specimen images. Consequently, the fixed images ($F$) consist of the manually registered histology images, whereas the ground truth labels ($\gamma$) include the original snapshot specimen images. The VoxelMorph network ($g_\theta(F, M)$) is trained by the loss function ($L$) to transform $M$ to $F$ using the predicted DDF ($\varphi$) in combination with the spatial transformer function. The modified model with an example of our data is illustrated in Figure \ref{supervised}.

\bigskip

In the supervised approach, the loss function was calculated by comparing the predicted registered image (M($\varphi$)) and the ground truth label image ($\gamma$), which both have the same modality, similar image intensity distributions, and local contrast. Therefore, the mean squared error ($MSE$) loss function is used as $L_{sim}$ in Equation \ref{equation1} which is described in Equation \ref{equation3} \cite{Balakrishnan2018AnRegistration}

\begin{equation}
\label{equation3}
    \text{MSE}(\gamma, M(\varphi)) = \frac{1}{n} \sum_{i=1}^n [\gamma_i - M(\varphi)_i]^2
\end{equation}

Where, $n$ is the total number of samples, $\gamma_n$ the ground truth image for the $n$-th sample (original snapshot specimen image), and $M(\varphi)_n$ the predicted value for the $n$-th sample (predicted registered image).

\subsubsection{Training}
\label{training}
We utilized Python (version 3.10.4) along with the TensorFlow \cite{AbadiTensorFlow:Systems} and Keras \cite{GitHubHumans} libraries for data manipulation and analysis. The augmented dataset was split into three subsets, whereas, 360 paired deformed snapshot specimen images will be allocated for the training set, 90 paired images for the validation set, and 115 paired images for the test set. To train the network, the ADAM [79] optimizer with a learning rate of 0.001 was used. The configuration involved setting the number of epochs to 200, with 100 steps per epoch and a batch size of 16.

\subsubsection{Evaluation matrices}
\label{evaluation}
In order to assess the performance of automatic deformable registration models described in Section \ref{unsupervised model} and \ref{supervised model}, various evaluation metrics were employed. This evaluation was carried out for all images in the test set, both before and after applying the registration models.
The Dice score was used to measure the degree of overlap between two binary images ($A$ and $B$) by comparing the number of common pixels in the two images with the total number of pixels in the reference image ($B$), as described in Equation \ref{equation4}. This metric is especially used to evaluate the overlap of the boundaries of the images.

\begin{equation}
\label{equation4}
\text{Dice}(A,B) = 2\frac{|A\cap B| }{|A|+|B|}
\end{equation}

The histogram-based mutual information ($HMI$) was used to measure the similarity between images by comparing their histograms (Equation \ref{equation2}). The $HMI$ between two images is the amount of information that is shared between their histograms. Specifically, it measures how much the joint histogram of the two images deviates from the product of their individual histograms. As a result, the optimal alignment of the two images can be determined.

%At last, the target registration (TRE) is computed as showed in Equation ... . To perform this evaluation, ... random image pairs from the test set were selected. Next, landmark pairs were chosen on both the moving image ($P_m$) and the ground truth image ($P_f$). The points from the moving image were transformed using the predicted DDF. The coordinates of these transformed points were then compared with the coordinates of the selected ground truth points. The TRE was calculated as the normalized distance between the landmarks of the ground truth ($P_f$) and the transformed landmarks of the moving image ($T(P_m)$).

%\begin{equation}
%\label{equation5}
%    \text{TRE} = ||T(P_m) - P_g||
%\end{equation}

For both obtained Dice and $HMI$ metrics, statistical analysis was performed between the unregistered and registered results using IBM SPSS statistics v27 (SPSS Inc., United States). %Normal distribution was assessed with the Shapiro-Wilk test. 
Statistical analysis for non-normally distributed data was performed using a Mann-Whitney test. Whereas, a p-value $\leq$0.05 was considered statistically significant. 

%%%%%%%%%%%%%%%%%%%%%%%%%%  Results  %%%%%%%%%%%%%%%%%%%%%%%%%%

\section{Results}

\subsection{Evaluation unsupervised and supervised models}
\begin{figure}[!h]
    \centering
    \includegraphics[width=0.95\textwidth]{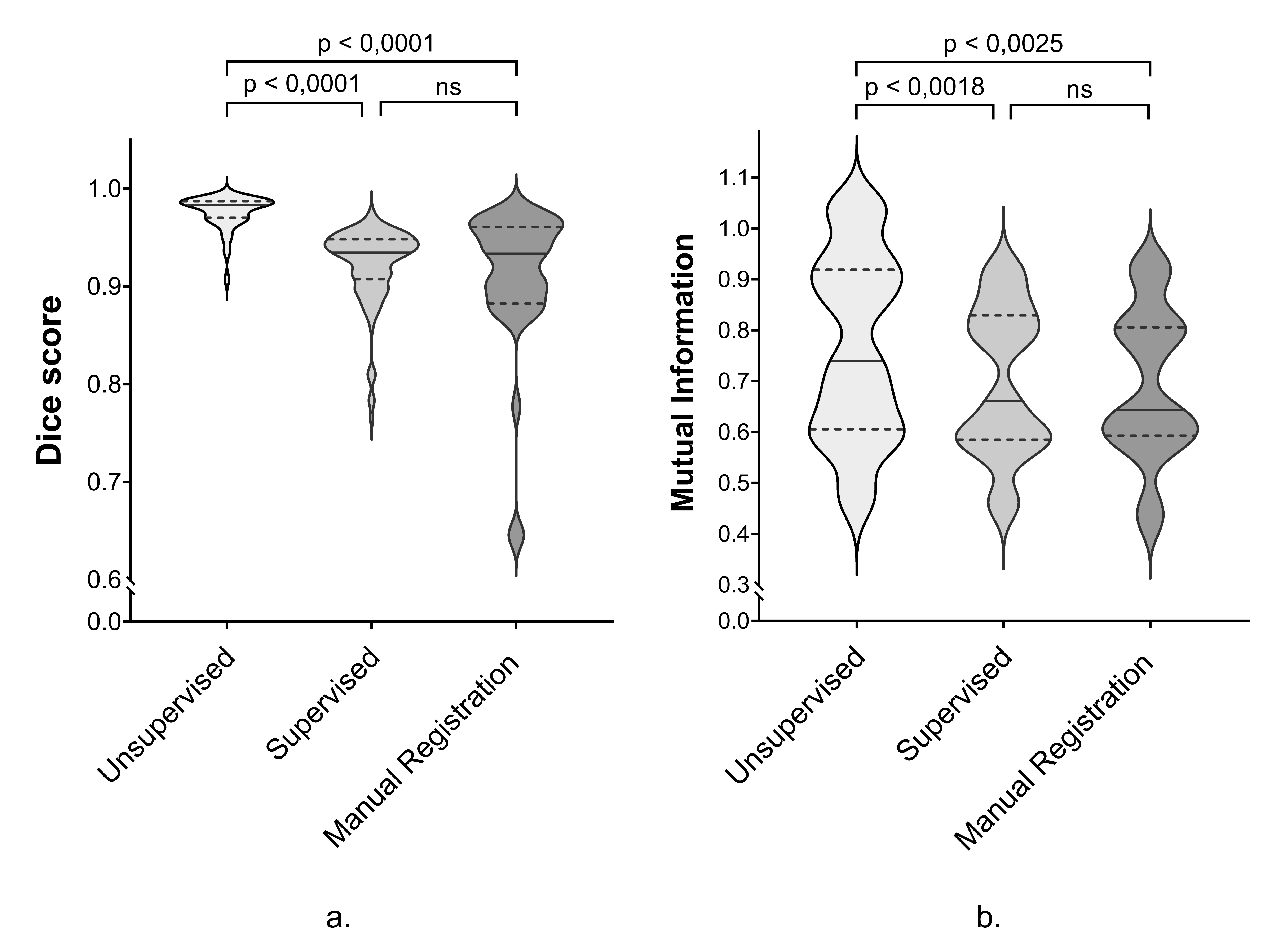}
    \caption{Evaluation automatic deformable image registration method for the unsupervised, supervised and manual approaches where (a) Dice score and (b) Mutual information values are displayed for 115 specimen pairs after the registration. The solid line represents the median, whereas the dashed lines represent the interquartile range (IQR).}
    \label{Results}
\end{figure}

Figure \ref{Results} visualizes the Dice score and mutual information for the results of unsupervised and supervised approaches compared to the manual registration. 

The violin plots in Figure \ref{Results} show the distribution of evaluation metrics for the same pair of specimen images from the test set. In case of the unsupervised and supervised approaches, metric values were calculated between the fixed ($F$) and predicted image (M($\varphi$)). For the manual registration, metric values between the manual registered histology image ($F$) and the specimen snapshot image (label $\gamma$) are reported. The width of these plots shows the relative frequency in which each value occurs, and becomes wider when the value occurs more frequently and with a higher probability. 

The distribution of Dice scores ranges from 0,90-0,99 (median 0,98 $\pm$ 0,02) and 0,77-0,92 (median 0,92 $\pm$ 0,04) for the unsupervised and supervised approaches, respectively. The mutual information for the unsupervised method is distributed in a range between 0,45-1,00 (median 0,76 $\pm$ 0,18), which is slightly higher compared to the supervised method where the values are distributed between 0,44-0,94 (median 0,74 $\pm$ 0,13). 

\begin{figure}[!h]
    \centering
    \includegraphics[width=0.80\textwidth]{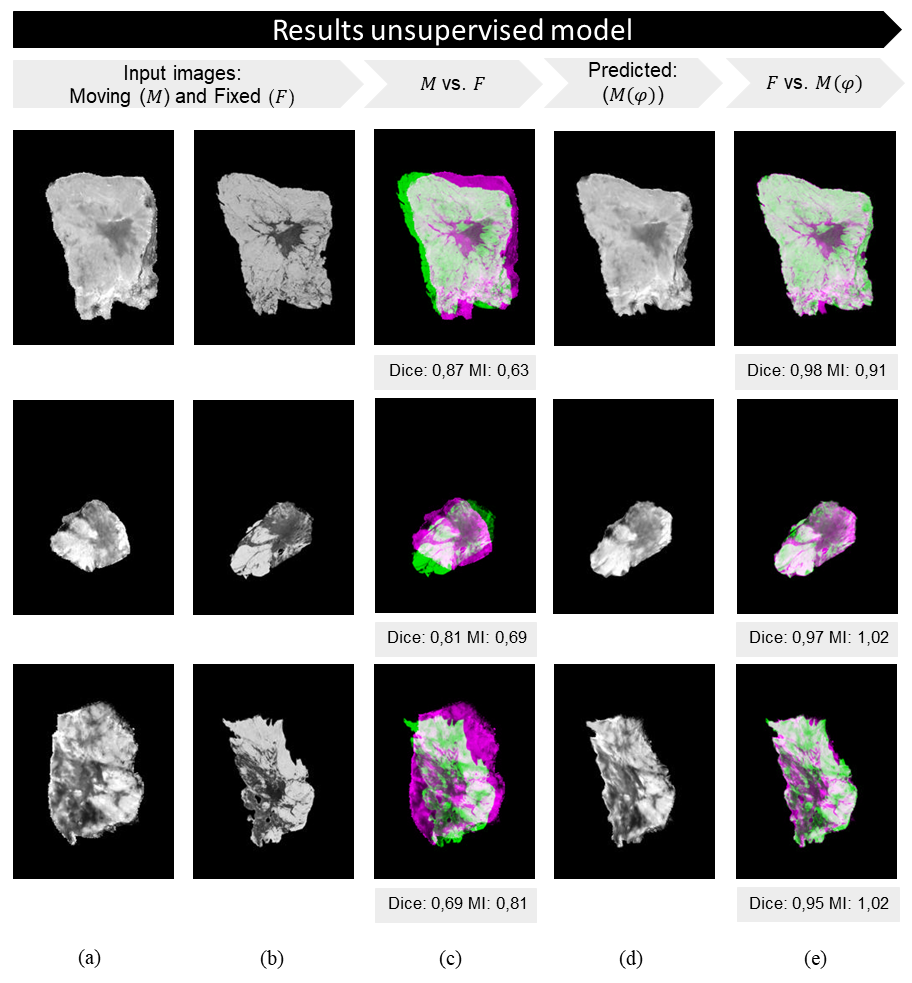}
    \caption{Results unsupervised model: a. Specimen snapshot image ($M$) b. Artificially deformed histology image ($F$) c. Unregistered images: overlap between $M$ and $F$ d. Predicted image $M(\varphi)$ e. Registered images: overlap between $F$ and $M(\varphi)$. Dice and mutual information are showed for the unregistered and registered examples.}
    \label{results unsupervised}
\end{figure}

\begin{figure}[!h]
    \centering
    \includegraphics[width=0.80\textwidth]{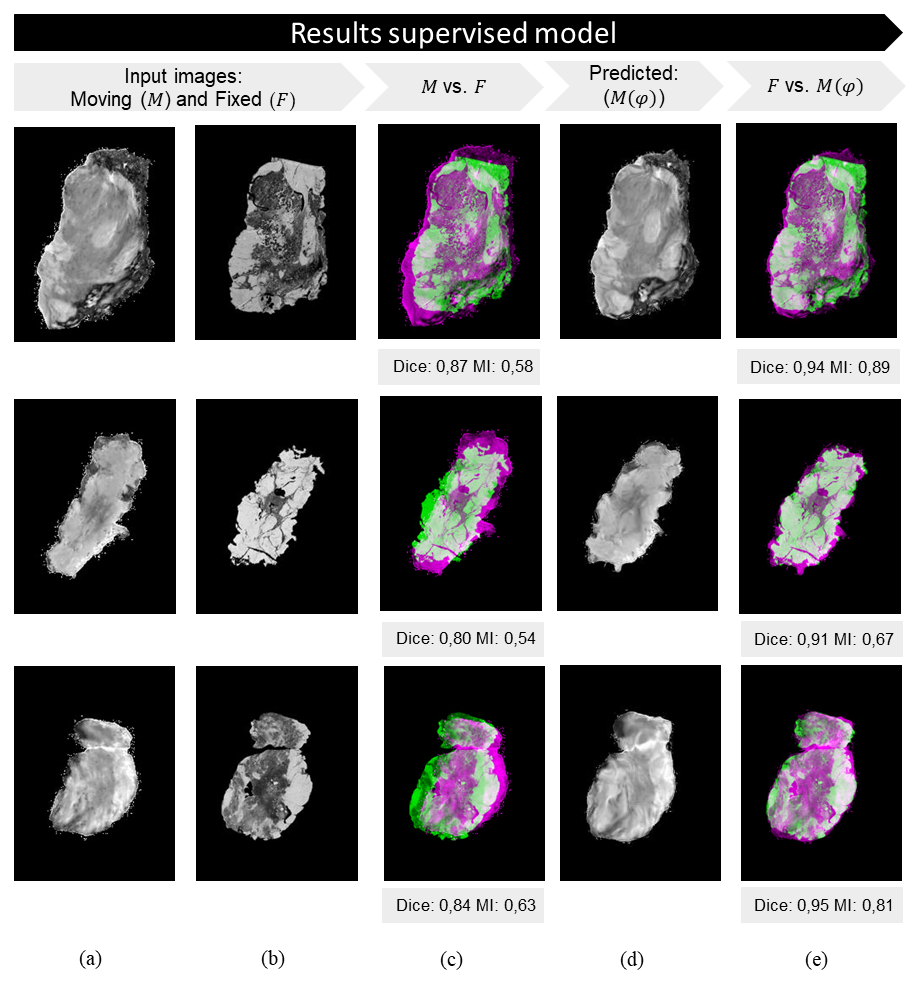}
    \caption{Results supervised model: a. Artificially deformed specimen snapshot image ($M$) b. The manual registered histology image ($F$) c. Unregistered images: overlap between $M$ and $F$ d. Predicted image $M(\varphi)$ e. Registered images: overlap between $F$ and $M(\varphi)$. Dice and mutual information are showed for the unregistered and registered examples.}
    \label{results supervised}
\end{figure}

Figures \ref{results unsupervised} and \ref{results supervised} display multiple registration examples from the test set for the unsupervised and supervised approach applied on the same paired specimen images.

%%%%%%%%%%%%%%%%%%%%%%%%%%  Discussion  %%%%%%%%%%%%%%%%%%%%%%%%%%

\section{Discussion}
%Introduction: Summary difficulties multi-modal registration, validation of optical technologies, correct labeling important and goal paper + potential benefits. 
When achieving a precise registration between the optically measured tissue and histopathology, the development of tissue classification algorithms can be optimized, which thereby improves the effectivity of optical technologies in clinical practice. 
However, registration difficulties arise when dealing with deformed multi-modality images, such as histology and tissue specimen images. Utilizing sophisticated algorithms and computational methods to optimize deformable registration processes holds promise in overcoming current inaccuracies in the validation of optical technologies. In this paper, we explored unsupervised and supervised implementations based on the VoxelMorph model, to achieve a deformable registration between 2D multi-modal images. We used a previously in-house acquired dataset of manually registered images breast specimen images to train the models. 

%%%%%% Discussing results %%%%%%% 

%unsupervised, supervised and comparison between techniques. 
The efficacy of the developed models was assessed through the computation of both Dice scores and mutual information for all 115 registered images, the overlap between $F$ and $M(\varphi)$, in the test set (Figure \ref{Results}). The unsupervised method outperformed the other approaches significantly. Specifically, as indicated by the Dice score, a more accurate overlap between the general shape of the input images was achieved. Mutual information (MI) functioned as a metric for assessing the similarity between distinct image modalities. As illustrated by the violin plot, the unsupervised dataset, demonstrated a prominently increased distribution within the 0.6-1.0 range, indicating an improved alignment of internal structures compared to the other approaches. 
Unlike mono-modal image registration, where the ground truth transformation is available, multi-modal images do not have a direct one-to-one correspondence due to differences in imaging modalities. This makes it difficult to define an objective reference for evaluating the accuracy of registration. The dataset used in this study is unique since it contains ground truth manually registered histology images, which are not commonly available in similar datasets. This makes the dataset particularly well-suited for developing and testing multi-modal image registration algorithms and other image analysis techniques.
However, training a model in a supervised manner, with labels derived from manually registered ground truth images, showed only a slight improvement in comparison to the manual registration approach. This can be explained due to the fact that this model was trained with images which possibly contain small manual registration errors. Besides the use of labels, the main difference between the training of supervised and unsupervised models involves the used loss function ($L$). Specifically, for the purpose of this study, the mutual information-based loss function demonstrated superior performance compared to the Mean Squared Error (MSE).
 
%%%%%% Comparing with other literature %%%%%%%%
The adoption of algorithms for automating the deformable registration processes represents a paradigm shift in image registration, offering distinct advantages over manual methods \cite{Lu2014HyperspectralAccess,Halicek2018DeformableDemons,DeBoer2019MethodDeformations}. Our results demonstrate that the unsupervised algorithm achieved superior significant performance when compared with the ground truth manual point-based registration, which emphasizes the use of applicability of computational techniques for multi-modal image registration. In comparison, manual registration methods, and corresponding pre-processing steps, can be prone to human errors and inconsistencies, making the automated approach a significantly more reliable option. Recently, there has been a growing acknowledgment among studies regarding the essential requirement to account for tissue deformations when correlating optical measurements with a ground truth pathology label \cite{Zhang2021ExplainableLearning, Phipps2018AutomatedImaging}. Multi-modal registration is often complicated by the lack of corresponding landmarks between images. Therefore, the use of fiducial markers is investigated but involves invasive procedures, such as the placement of burn marks on the tissue surface, that could potentially inflict damage on delicate tissue structures \cite{Unger2018MethodAssessment}. Besides, manual tasks are characterized by their labor-intensive nature, which demand considerable time investments to ensure accurate alignment. In contrast, the inherent efficiency of automatic registration accelerates the alignment process, minimizing the potential for discrepancies and enhancing the overall quality of results.

%%%%%% Limitations & Recomandations %%%%%%% 

% suitable evaluation metrics, mutual information difficult to interpreter. MI also dependent on the used pre-prosessing. 
% Creating landmarks and calculate TRE.
While advancements in registration algorithms have significantly improved the accuracy and robustness of image alignment, the selection of appropriate evaluation metrics remains a challenging and nuanced task. The complexities inherent to multi-modal registration pose a range of difficulties in identifying evaluation metrics that accurately assess the quality of registration outcomes. Multi-modal registration often involves non-linear transformations to account for differences in anatomical structures and intensities across modalities. Conventional metrics such as mean squared error or mutual information, which are effective for linear transformations, may inadequately capture the intricate deformations and intensity variations inherent to multi-modal registration. Our findings indicate precise registration that effectively compensates for deformation, even for the visible internal structures. However, this achievement is associated with relatively low Mutual Information (MI) values, explicable by the inherent variations in contrast among multi-modal images and the employed preprocessing procedures. The challenge lies in devising metrics that can appropriately quantify the alignment accuracy across diverse spatial and intensity changes. Additional metrics like target registration error could be considered to provide a more conclusive assessment of the model's performance.

%Holes in HE images complicates registration model. holes in HE make registration difficult, but not related to performance model.
The complex task of registering microscopic histology images with their corresponding tissue slices in RGB encounters challenges arising from the fundamental differences between these imaging modalities. Microscopic histology images, revealing details at the cellular level, are typically acquired through staining and specialized imaging techniques. In contrast, RGB images offer a macroscopic perspective of tissue slices under conventional optics, capturing color information at a higher scale. The presence of tears and holes disrupts the natural continuity of cellular structures in histology images, introducing gaps and inconsistencies that challenge the registration process. Developed registration techniques therefore struggle to establish trustworthy correspondences between regions that are  distorted by these artifacts. Tears introduce non-local deformations, while holes disrupt the continuity of anatomical features, making it challenging for algorithms to accurately match corresponding areas in RGB images. Therefore, it is essential to acknowledge that suboptimal performance of this developed model can, at times, be influenced by the degree of deformation and the presence of artifacts in the histology images.

%Train models for different applications. 
The presented approach has the potential to optimize the registration efficiency, for breast tissue specifically, ultimately leading to an enhancement in the precision of correlating optical measurements with a correct pathology label used for the development of tissue classification algorithms. Further research should also focus on exploring the suitability of this developed model for deformation problems in histology images which occur across different tissue types. This can potentially result in an improved optical technology validation for multiple organ domains, which ensures the dependable integration of optical technologies into clinical practice.

%%%%%%%%%%%%%%%%%%%%%%%%%%  Conclusion  %%%%%%%%%%%%%%%%%%%%%%%%%%

\section{Conclusion}
In conclusion, our efforts have resulted in the development of an automated multi-modal image registration technique based on deep learning principles. This method effectively aligns snapshot breast specimen images with corresponding histology images, achieving a high degree of precision. Notably, the performance of the unsupervised model exceeds that of a previous manual approach, presenting a faster and significantly more accurate registration method. This advancement holds the promise of improving the validation of optical technologies across diverse organ domains, ensuring the reliable integration of optical tools into clinical practice.

\section*{Funding} The authors gratefully acknowledge the financial support of this research by the Dutch Cancer Society (Grant No. KWF 13443).

\section*{Acknowledgements} The authors would like to thank all surgeons and nurses from the Department of Surgery, all pathologist assistants from the Department of Pathology for their assistance in processing specimens, the NKI-AVL core Facility Molecular Pathology \& Biobanking (CFMPB) for supplying NKI-AVL biobank material and all students that participated in this research for their time and effort. Research at the Netherlands Cancer Institute is supported by institutional grants of the Dutch Cancer Society and of the Dutch Ministry of Health, Welfare and Sport.

\bibliographystyle{unsrt}
\bibliography{references.bib}

\end{document}